\documentclass{revtex4}
\usepackage{amsmath}
\usepackage{graphicx}
\usepackage{float}
\usepackage{wrapfig}
\usepackage{subfig}

\begin{document}

\title{Effects of magnetic drifts on ion transport in turbulent tokamak plasmas}
\author{A. Croitoru}
\affiliation{National Institute of Laser, Plasma and Radiation Physics,
	PO Box MG 36, RO-077125 M\u{a}gurele, Bucharest, Romania}
\affiliation{Faculty of Physics, University of Bucharest,
	PO Box MG 11, RO-077125 M\u{a}gurele, Bucharest, Romania}

\begin{abstract}
The stochastic advection of low energy deuterium ions is studied in a three dimensional realistic turbulence model in conditions relevant for current tokamak fusion experiments. The diffusion coefficients are calculated starting from the test particles trajectories in the framework of the semi-analitical statistical model called the Decorrelation Trajectory Method. We show that trajectory trapping determined by the space correlation of the velocity field is the main cause for anomalous diffusion and we obtain the transport regimes corresponding to different values of the parameters of the turbulence model and a trapping condition in the limit of frozen turbulence. Depending on the value of the parameters of the model, the interaction between turbulence and magnetic drifts leads to both an increase (with transport in the poloidal direction increasing up to an order of magnitude) and a decrease of the particle transport. 

\end{abstract}

\maketitle

\section{Introduction}

The efficiency of magnetic fusion experiments is conditioned by the rate of transport of thermal energy and particles from the hot core of the plasma to the colder edge. The effective transport coeficients obtained in tokamaks by means of particle and power balance studies have a much larger value than what would be expected solely from collisional neoclassical transport
processes \cite{hasegawa1987self}. This anomalous transport is attributed in large part to the presence of turbulent processes, such as drift waves instabilities excited by radial gradients in density and in temperature. The field is very active and there exist a vast literature dedicated to the study of the effect of various turbulence scenarios as for exemple the Ion Temperature Gradient (ITG) mode and the Trapped Electron Mode (TEM) instabilities on transport, with recent gyrokinetic simulations including also the coupling ITG-TEM \cite{qi2016gyrokinetic} or the dissipative
trapped electron mode (DTEM) that causes electrostatic turbulence in the pedestal \cite{zhao2017gyrokinetic}.

While the behaviour of energetic ions has reached some satisfactory level of understanding, with studies offering scallings of transport with microturbulence [\cite{hauff2008mechanisms, hauff2010scaling, zhang2010scalings}, in both 2D \cite{hauff2007b} and 3D turbulence \cite{hauff2008mechanisms} low energy ions have received little to no attention despite studies suggesting that the diffusivity of lower energy test particles is similar to that of fast particles \cite{gunter2007interaction}. 

In the present paper we study the collisionless turbulent advection of low energy deuterium ions in a three-dimensional tokamak geometry and in a realistic turbulence model corresponding to the general characteristics of ITG or TEM, using a semi-analitical test particle approach  developed by M.Vlad et al. \cite{vlad1998diffusion, vlad2013test} called The Decorrelation Trajectory method (DTM). We show that the interaction between the neoclassical transport processes and the turbulent processes in the presence of trapping (for high values of the Kubo number) is highly nonlinear and can lead to either an increase or to a decrease of the diffusion coefficients depending on the parameters of the turbulence model. The spectrum of potential fluctuations is modeled as in \cite{croitoru2017turbulent} to be in agreement with the results of the numerical simulations \cite{shafer20122d, hauff2007b} and includes the drift of the potential with the effective diamagnetic velocity and parallel decorrelations. Depending on the value of its velocity pitch-angle $\theta$ a particle in a three dimensional nonuniform magnetic field is either passing or trapped. We show that the value of the radial diffusion coefficient is maximum for $\theta=0$.

The paper is organized as follows. In Sec. \ref{turbulencemodel}  we describe the turbulence model. Sec. \ref{statisticalmethod} offers a brief description of the DTM applied to particles moving in a 3D turbulent electromagnetic field. The results of the interaction between the drift of the turbulent potential and the magnetic drift are summerized in Sec. \ref{results} and the conclusions in Sec. \ref{conclusions}.

\section{Turbulence Model}\label{turbulencemodel}

When the distance covered by the particles in a time length of the order of the decorrelation time is much smaller than the gradient scale length of the nonuniform quantities (such as temperature or density) we can apply a test particle approach \cite{balescu2005aspects}. 
Unlike in self-consistent models, the turbulence model in the test particle method is considered given a priori as a function of some turbulence parameters and independent of the distribution function of the particles. This allows us to obtain the transport coefficients as functions of the Fourier transform of the spectrum of the electrostatic potential (Eulerian correlation), leading to different transport regimes corresponding to different ranges of the parameters of the turbulence model.  

The electrostatic potential of a plasma in a drift type turbulence has an irregular structure of hills and wells continuosly changing in time \cite{vlad2004lagrangian}. This is mainly due to 
the presence of density gradients which drive diamagnetic currents that restore the equilibrium. If there exists a small disturbance in the particles' pressure gradient, the response of the diamagnetic currents is also perturbed and it is propagated in a direction perpendicular to the magnetic field by ions that move with the ion polarization drift velocity \cite{horton1999drift}. This then drives a perturbed parallel current of the electrons so that the total current is divergence-free. If the parallel electron motion is not adiabatic and dissipates because of the interaction with the background plasma, there will be a time delay between the electron density perturbation and the plasma potential perturbation leading to a growing amplitude of the disturbance. The resulting nonzero time average product of the density and the $\mathbf{E}\times \mathbf{B}$ velocity causes a net transport of the plasma with an effective diamagnetic velocity wich can be either in the direction of the electron or of the ion diamagnetic velocity. 
The growing drift wave disturbance can interact nonlinearly with disturbances at other wavenumbers leading to perturbations at other scales, either at smaller wavenumbers due to kinetic energy perturbations or at larger wavenumbers due to density perturbations.
The large scale perturbations, as for example the zonal flow, decorrelates and fragmentates the initial structure of the potential dramatically modifiyng the trasport.
It follows that the spectrum of the fluctuating potential must contain two separated peaks \cite{balescu2005aspects}, one corresponding to the large scale oscillations and the other to the small scale ones. This aspect is confirmed in numerical simulations such as \cite{shafer20122d, hauff2007b} that show a spectrum
with two symmetrical maxima at $k_y=\pm k_y^0$ and $k_x=0$ and with a vanishing amplitude for $k_y=0$. Since ion dynamics in the poloidal direction is responsible for the potential fluctuation, the spectrum at saturation will have a similar form for both ITG and TEM instabilities, with the only difference being the typical values of the poloidal wave numbers ($k_y\rho_s\sim 1$ for TEM and $k_y\rho_s \sim 0.1-0.5$ for ITG) \cite{doyle2007plasma}.
We can thus approximate this spectrum by the analytical expression \cite{croitoru2017turbulent}
\begin{equation}
\begin{split}
S(k_{1},k_{2},z,t) &\propto A_\phi ^{2} \left[ \exp \left( -\frac{(k_{2}-k_{2}^{0})^{2}}{2}\lambda
_{2}^{2}\right) -\exp \left( -\frac{(k_{2}+k_{2}^{0})^{2}}{2}\lambda
_{2}^{2}\right) \right]   \label{spectr} \\
& \times\frac{k_{2}}{k_{2}^{0}}\exp \left( -\frac{k_{1}^{2}}{2}\lambda
_{1}^{2}\right) \exp \left( -\frac{|z|}{\lambda _{z}}-%
\frac{|t|}{\tau _{c}}\right),  
\end{split}
\end{equation}
where we also included the decorrelation of the potential due to its variation in time and to the existence of a finite parallel decorrelation length $\lambda_z$.

\begin{figure}
  \centering
  \includegraphics[width=.45\linewidth]{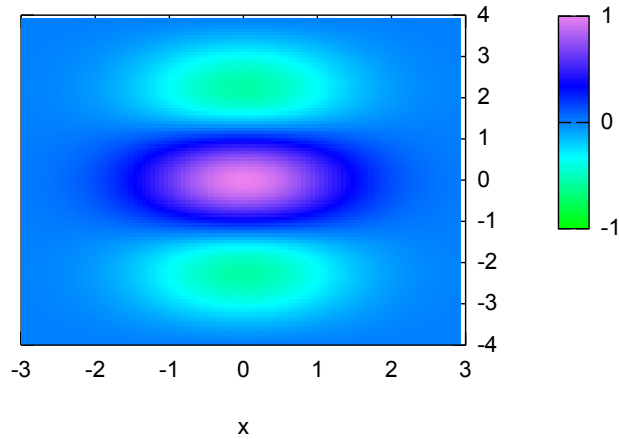}
  \caption{Normalized Eulerian correlation function of the porential $E(\mathbf{x},z,t)$}
  \label{spectru}
\end{figure}



The normalized Eulerian correlation (EC) of the potential (see figure Fig. \ref{spectru}) is obtained as the Fourier
transform of $S(k_{1},k_{2},z,t)$ :

\begin{equation}
\begin{split}
E(\mathbf{x},z,t) & =A_\phi^2\exp\left(-\frac{x^2}{2\lambda_1^2}\right)\cdot\frac{d}{dy}\left[\exp\left(-\frac{y'^2}{2\lambda_2^2}\right)\frac{\sin(k_2^0y')}{k_2^0}\right]\\
& \times\exp\left(-\frac{v_\parallel t}{\lambda_z}-\frac{t}{\tau_c}\right)
\end{split}
\end{equation}

The potential $\phi(x,y',z,t)=\phi(x,y-V_dt,z,t)$is modeled as a stationary and homogeneous Gaussian stochastic function. The potential drift with the diamagnetic velocity $V_d$ in the poloidal direction, specific to the drift type turbulence, is included in the argument $y'=y-V_dt$. 

The turbulence parameters are therefore the amplitude of the potential
fluctuations $A_\phi ,$ the correlation lengths along each direction $\lambda
_{i},$ $i=x$ (radial), $i=y$ (poloidal), $i=z$ (parallel), the correlation
time $\tau _{c}$ the dominant wave number $k_{2}^{0}$ and the average diamagnetic velocity $V_d$.

The normalized 3D equations of motion in a nonuniform magnetic field $\vec{B}=B_0e^{-x/L_b}\vec{e}_z\approx B_0\left(1-x/L_b\right)$ in the guiding center approximation in the frame of the moving potential are 

\begin{equation}
\begin{split}\label{ecdxdt}
\frac{dx_i}{dt}&=-K_*\epsilon_{ij}\partial_j\phi\left(1+x*\delta b\right)+\delta_{iy}V_d + \delta_{iy}\frac{\delta b}{\rho_*}\left(v_\parallel^2+\frac{v_\perp^2}{2}\right) ,\\
\frac{dz}{dt}&=v_\parallel, 
\end{split}
\end{equation}
$\delta b=v_\perp/(\omega_c*L_b)$ is the inverse of the gradient scale-length of the magnetic field $L_b$ normalized to the Larmor radius of the deuterium ions $\rho=v_\perp/\omega_c$, $\omega_c = qB_0/M_D$ is the cyclotron frequency and $\rho_*=\rho/a$, where $a$ is the small radius of the tokamak.   
The last term of the perpendicular velocity is the neoclassical gradient and curvature drift which is modelled as an average velocity with a Maxwellian distribution.
The following units were used: the Larmor radius $\rho$ for the  perpendicular displacements, the small radius of the tokamak $a$ for the parallel displacements, the thermal velocity $v_{th}$ for the neoclassical velocity, the thermal velocity times the normalized Larmor radius $v_*=\rho_* v_{th}$ for the diamagnetic velocity and the time needed for a particle movind radially with the thermal velocity to escape to the wall of the tokamak $\tau_0=a/v_{th}$ for time.
The dimensionless parameter
\begin{equation}
K_*=\frac{A_\phi}{B_0\rho}\frac{1}{v_*}=\frac{eA_\phi}{T}\frac{a}{\rho}=\frac{eA_\phi}{T}\frac{1}{\rho_*}
\end{equation}
is a measure of the amplitude of the turbulence.

\section{Statistical Method}\label{statisticalmethod}

The decorrelation trajectory method was developed by M.Vlad et al. \cite{vlad1998diffusion, vlad2017random, vlad2015nonlinear, vlad2006impurity} for the study of diffusion in incompresible stochastic velocity fields
\begin{equation}\label{demo}
\frac{dx_i}{dt}=K v_i(\mathbf{x},t)
\end{equation}
beyond the quasilinear regime, i.e at high values of the Kubo number $K$, where trajectory trapping or eddying determines anomalous statistics such as non-Gaussian distributions and memory effects expressed as long time Lagrangian correlations and also an increased degree of coherence \cite{vlad2017random}.  Defined as the ratio between the decorrelation time of the trajectories from the potential, $\tau_c$ and the time of flight of the particles, $\tau_{fl}=\lambda_c/V$ , where $\lambda_c$ is the space scale of the potential and $V$ is the amplitude of the fluctuating velocity field, the Kubo number is a measure of the trapping of the particles in the fluctuating potential as well as a measure of the amplitude of turbulence. As emphasized in \cite{vlad2017random} for $K\to\infty$ the particles are trapped indefinetly in the structure of the electrostatic potential, and partially for $K>1$. Perturbations to the Hamiltonian structure of the system introduce additional characterstic decorrelation times, such as the parallel decorrelation time $\tau_z=\lambda_z/v_z$.
As shown by Taylor \cite{taylor1922diffusion} the running diffusion coefficients are calculated as time integrals of the Lagrangian autocorrelation function of the velocity 
\begin{equation}
D_{ii}(t)=\int_0^t d\tau L_{ii}(\tau), 
\end{equation}
where $L_{ii}(\tau)=\langle v_i(\mathbf{x_1},t_1)v_i(\tau;\mathbf{x_1},t_1) \rangle$, with $v_i(\tau;\mathbf{x_1},t_1)$ being the velocity of the particle at time $\tau$ calculted along the trajectory starting at $\mathbf{x_1}, t_1$.

The main idea of the decorrelation trajectory method is to obtain the diffusion coefficients by projecting the Langevin equation (\ref{demo}) in subensembles S of realisations of the stochastic field, determined by fixed values of the stochastic potential and velocity in the origin of the trajectories $\mathbf{x}=\mathbf{0}, t=0$:

\begin{equation}
(S):\quad \phi^0=\phi(\mathbf{0},0)\quad v^0=v(\mathbf{0},0).
\end{equation}

The $\mathbf{E}\times \mathbf{B}$ velocity is still Gaussian in a subensemble, but it's subensemble average $\langle \mathbf{v}(\mathbf{x}(t),t)\rangle^S$ is usually nonzero. This allows for the definition of an average trajectory in a subensemble by 

\begin{equation}
\frac{d}{dt}\langle \mathbf{x}(t)\rangle^S=\langle \mathbf{v}(\mathbf{x}(t),t)\rangle^S.
\end{equation}

Since the trajectories in a subensemble are overdetermined by the initial conditions ($\phi(\mathbf{0},0)$ and its derivatives), they are very similar.

We can thus replace the trajectories in a subensemble with a single trajectory, called the decorrelation trajectory, obtained as a time integral of the subensemble average of the Eulerian velocity calculated along this average trajectory

\begin{equation}
\frac{d}{dt}\mathbf{X}^S=\langle \mathbf{v}(\mathbf{X}^S(t),t)\rangle^S.
\end{equation}

The average velocity is obtained from the derivatives of the potential average in a subensemble

\begin{equation}
\langle v_i(\mathbf{x}(t),t)\rangle^S=-\epsilon_{ij}\partial_j\langle\phi(\mathbf{x}(t),t)\rangle^S,
\end{equation}
which, if we perform a change of basis from $\left(\partial_x \phi^0(\mathbf{x}(t),t), \partial_y \phi^0(\mathbf{x}(t),t) \right)$\\ to $ \left(| \nabla \phi^0(\mathbf{x}(t),t)|, \beta^0\right)$, where $\beta^0$ is the angle between $\partial_y \phi^0(\mathbf{x}(t),t)$ and the gradient of the potential and integrate over the modulus of the gradient, can be written as a function of the derivatives of the Eulerian autocorrelation of the potential:

\begin{equation}
\langle\phi(\mathbf{x}(t),t)\rangle^{S'}=\phi^0\frac{E(\mathbf{x},z,t)}{E(\mathbf{0},0,0)}+\sqrt{\frac{8}{\pi}} \cos\beta^0\frac{\partial_y E(\mathbf{x},z,t)}{V_1}-\sqrt{\frac{8}{\pi}} \sin\beta^0\frac{\partial_x E(\mathbf{x},z,t)}{V_2},
\end{equation}
with $V_1=\partial_{yy}E(\mathbf{0},0,0)$ and $V_2=\partial_{xx}E(\mathbf{0},0,0)$, where the new subensemble $S'$ is now defined by the values of $\phi^0$ and $\beta$ in the origin. 
Thus, the diffusion coefficient is 
\begin{equation}
D_{ii}(t)=\frac{V_i}{2\pi}\sqrt{\frac{\pi}{2}}\int_{-\infty}^\infty d\phi^0P(\phi^0)\int_0^{2\pi} d\beta^0\binom{\cos\beta^0}{\sin\beta^0}X_i^{S'},
\end{equation}
where $\phi^0$ is Gaussian $P(\phi^0)=\exp(-(\phi^0)^2/2)$.

For system (\ref{ecdxdt}) we modeled the neoclassical velocity as an average velocity with a Maxwellian distribution function  $P(v_\parallel,v_\perp)=\left(M_D/2\pi k_BT\right)^{3/2}\exp\left(-(v_\parallel^2+v_\perp^2)/2\right)$. The decorrelation trajectories for subensemble $S'$ are therefore 

\begin{equation}\label{DTs'}
\frac{dX^{S'}_i}{dt}=-K_*\epsilon_{ij}\partial_j\langle\phi\rangle^{S'}\left(1+X^{S'}*\delta b\right)+\delta_{iy}V_d + \delta_{iy}\frac{\delta b}{\rho_*}\left(v_\parallel^2+\frac{v_\perp^2}{2}\right),
\end{equation}
where
\begin{equation}
\begin{split}
\langle\phi(\mathbf{x}(t),t)\rangle^{S'}&=\phi^0\frac{E(\mathbf{x},z,t)}{E(\mathbf{0},0,0)}+\sqrt{\frac{8}{\pi}} \cos\beta^0\frac{\partial_y E(\mathbf{x},z,t)}{V_1}-\sqrt{\frac{8}{\pi}} \sin\beta^0\frac{\partial_x E(\mathbf{x},z,t)}{V_2}\\
&+x*V_d+x*\frac{\delta b}{\rho_*}\left(v_\parallel^2+\frac{v_\perp^2}{2}\right),
\end{split}
\end{equation}
and $V_1=\partial_{yy}E(\mathbf{0},0,0)=(k_2^0)^2+3/\lambda_y^2$ and $V_2=\partial_{xx}E(\mathbf{0},0,0)=1/\lambda_x^2$.

The dimensionless parameter $K_*$ is very similar to the Kubo number since it can be written as $K_*=\tau_0 / (\rho / V) $, where $V$ is the amplitude of the $\mathbf{E}\times \mathbf{B}$ velocity fluctuations $V=(A_\phi/B_0)/\rho$ in the chosen normalization (see section \ref{turbulencemodel}).

Defining the pitch angle $\theta$ as the angle between the modulus of the neoclassical magnetic drift and the projection of the magnetic drift on the $x$ axis, such that $v_\parallel=v_{neo}^0\cos\theta$ and $v_\perp=v_{neo}^0\sin\theta$, the running diffusion coefficient as function of the picth angle will be 

\begin{equation}\label{D(t)}
D_{ii}(t;\theta)=\frac{V_i}{2\pi}\sqrt{\frac{\pi}{2}}\int_0^\infty dv_{neo}^0 P(v_{neo}^0) \int_{-\infty}^\infty d\phi^0P(\phi^0)\int_0^{2\pi} d\beta^0\binom{\cos\beta^0}{\sin\beta^0}X_i^{S'},
\end{equation} 
with $P(v_{neo}^0)=\exp\left(-(v_{neo}^0)^2/2)\right)$. 

\begin{figure}
\captionsetup{justification=raggedright}  
\begin{minipage}[b]{.48\textwidth}
\centering
\subfloat[ $V_d=0$]{\includegraphics[width=.4\linewidth]{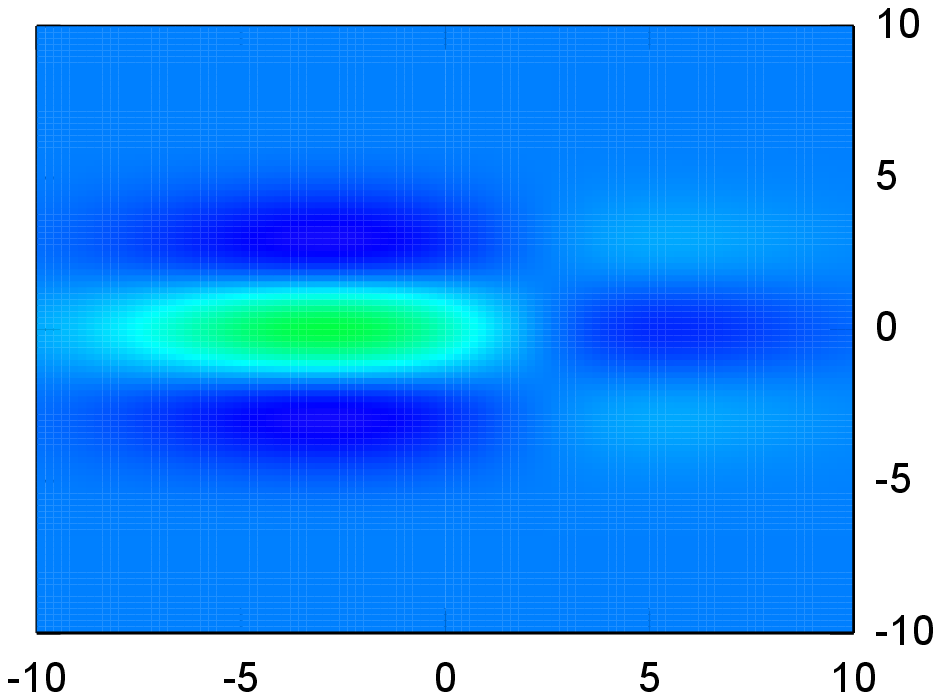}}
\hspace{0.4cm}
\subfloat[ $V_d=0.05$]{\includegraphics[width=.49\linewidth]{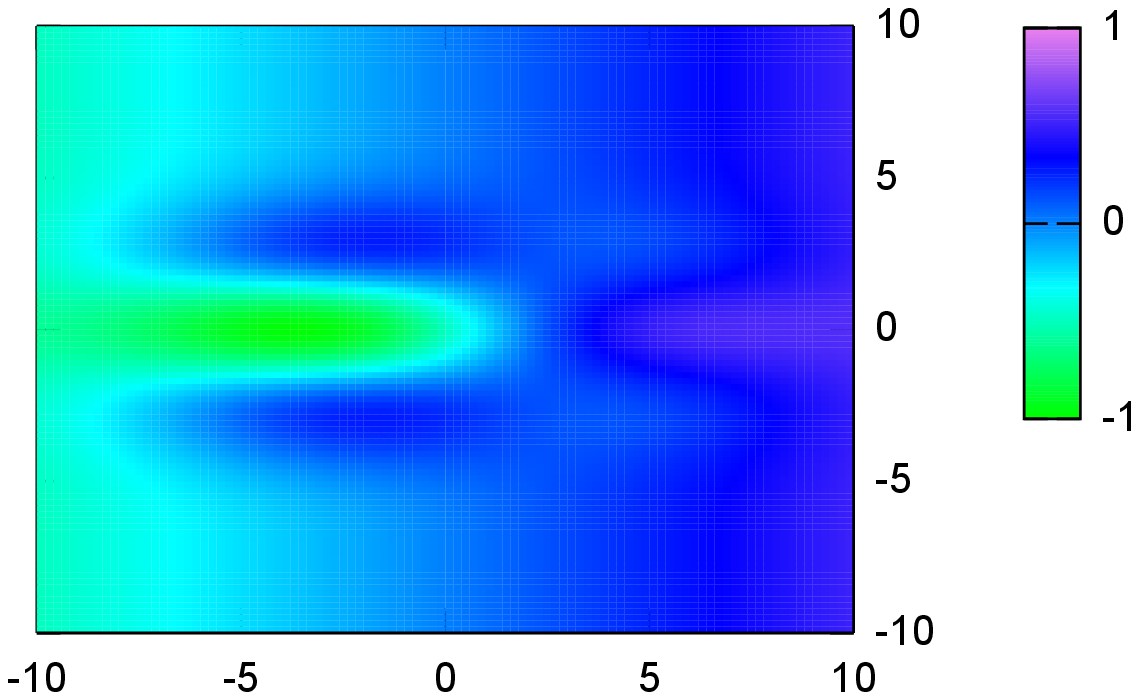}}\\
\subfloat[ $V_d=0.1$]{\includegraphics[width=.4\linewidth]{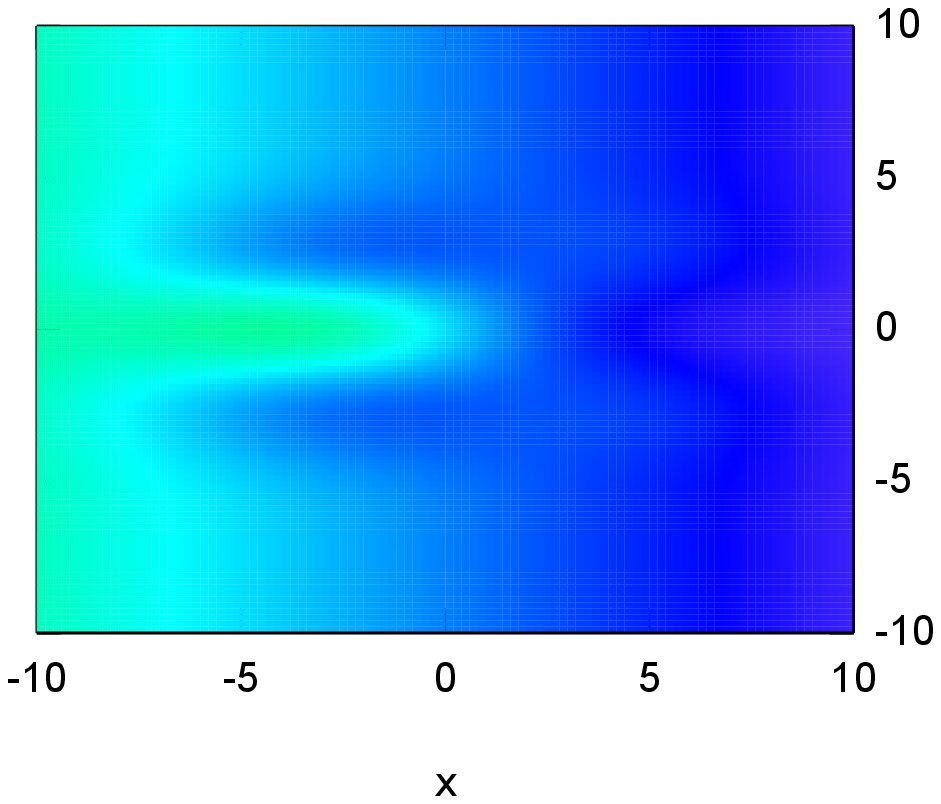}}
\hspace{0.4cm}
\subfloat[ $V_d=0.15$]{\includegraphics[width=.49\linewidth]{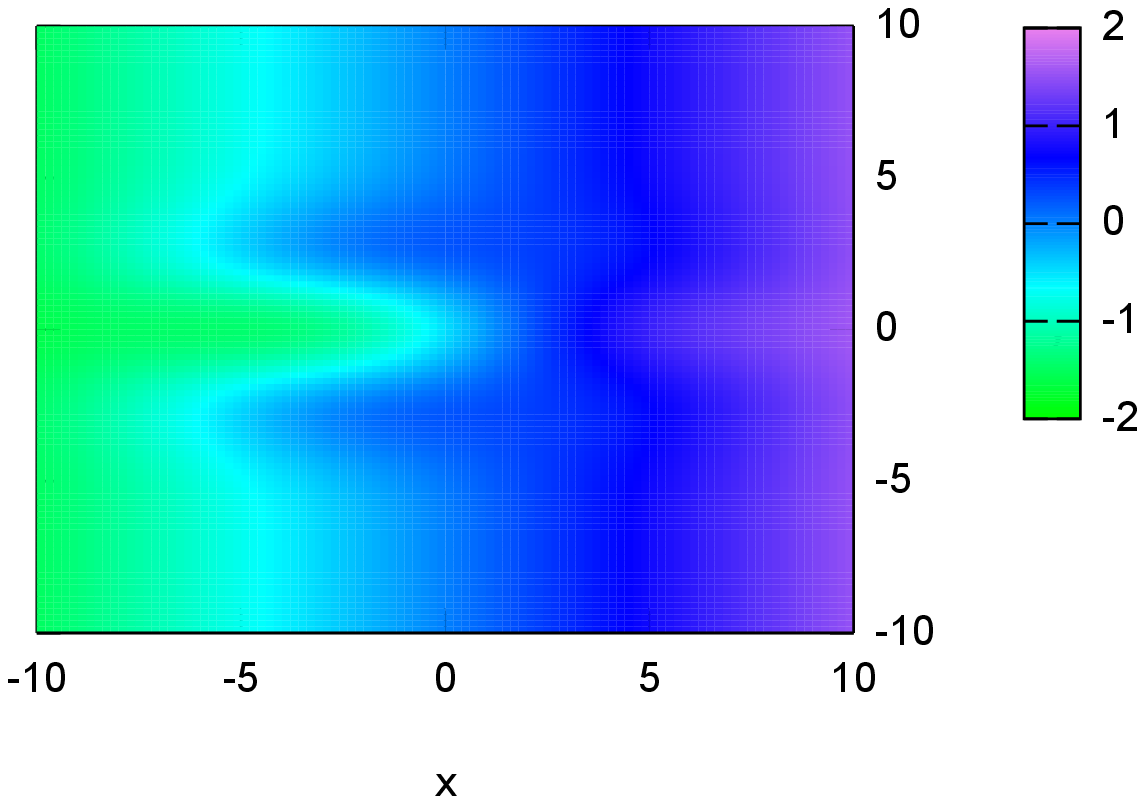}}\\
\caption{Normalized Eulerian Correlation of the potential at the plane $z=0, t=0$ for increasing values of the diamagnetic velocity}
\label{phimed}
\end{minipage}\hfill
\begin{minipage}[b]{.48\textwidth}
\centering
{\includegraphics[width=.99\linewidth]{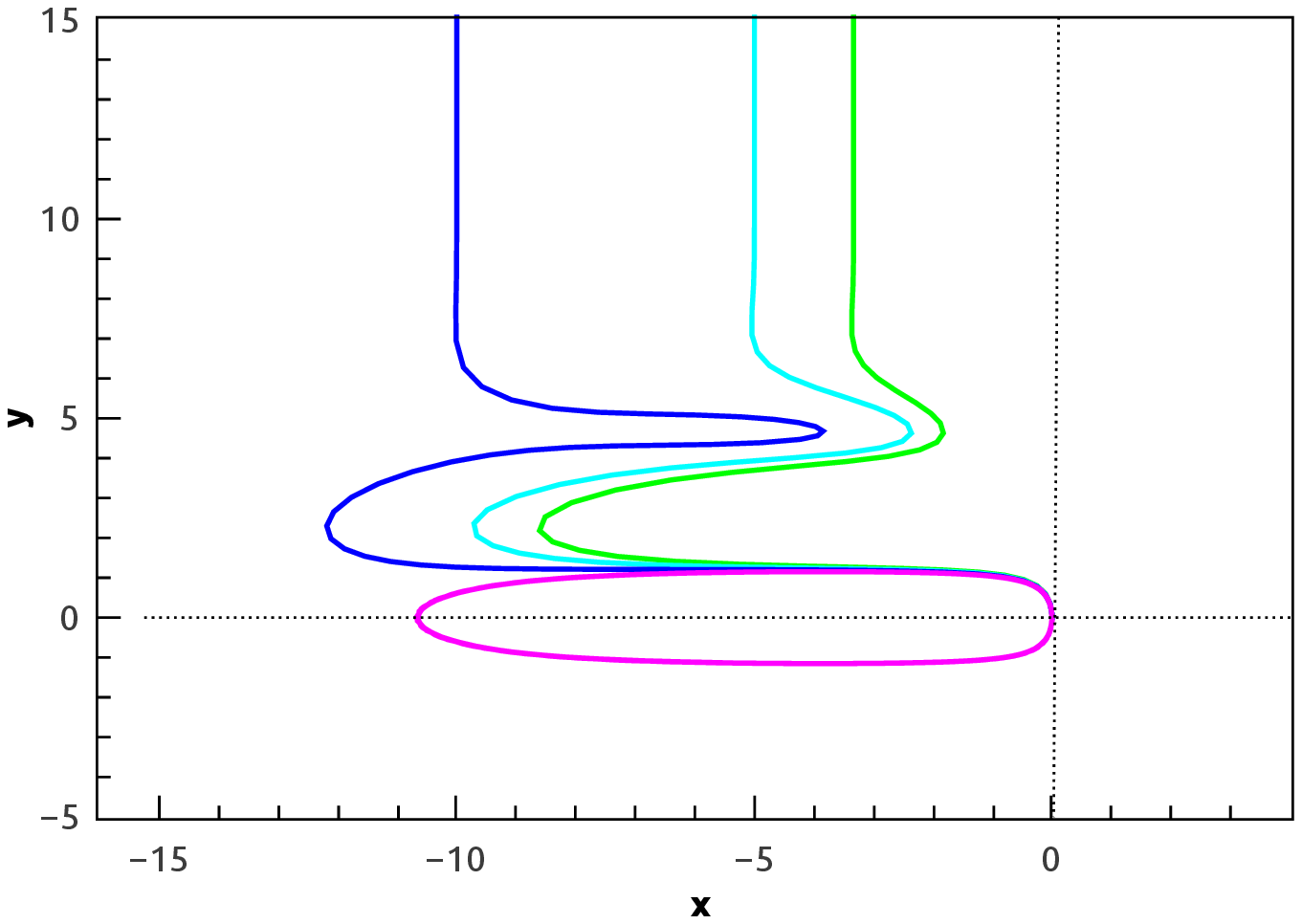}}\\
  \vspace{0.9cm}
\caption{Decorrelation trajectories corresponding to the (a)-(d) average potential(magenta $V_d=0$, blue $V_d=0.05$, ciel $V_d=0.1$, green $V_d=0.15$)}
\label{DTs}
\end{minipage}
\end{figure}

The running diffusion coefficients (\ref{D(t)}) are obtained numerically using approximately $40000$ test particles. System (\ref{DTs'}) is solved using an adaptive Runge Kutta method, the Cash-Karp method, implemented in the C++ boost library. The adaptive stepsize controls the error of the method and ensures stability. The trajectories are then interpolated at certain times using a spline interpolation technique and the integrals are calculated using the trapezoidal rule. 
In figure Fig. \ref{DTs} we plotted a decorrelation trajectory of a test particle starting at $x=y=0$ for increasing values of the diamagnetic velocity, in an constant magnetic field see Fig. \ref{phimed}. The drift of the potential with $V_d$ is equivalent to the existence of an average potential $xV_d$. This modifies the contour lines of the total potential. Since the magnetic drift is modelled as an average drift, it has the same effect as the diamagnetic velocity leading to an average potential $x*\frac{\delta b}{\rho_*}\left(v_\parallel^2+v_\perp^2/2\right)$. Their combined effect is to modify the contour lines of the total potential creating islands of closed contour lines with a decreasing size as the size of the drifts increases. At large $t$, the particles will move in the direction of this combined drift velocity, in this case, in the $y$ direction.

\section{Results}\label{results}
\begin{figure}
\begin{minipage}{.99\textwidth}
  \centering
  \subfloat[ $D_{xx}(t), D_{yy}(t)$ for different values of the normalized inverse gradient scale length (blue lines $\delta b=0$, green lines $\delta b=0.00041$) for a static potential $V_d=0$ with a parallel decorrelation length $\lambda_z=1$ ]{\includegraphics[width=.48\linewidth, height=.38\linewidth]{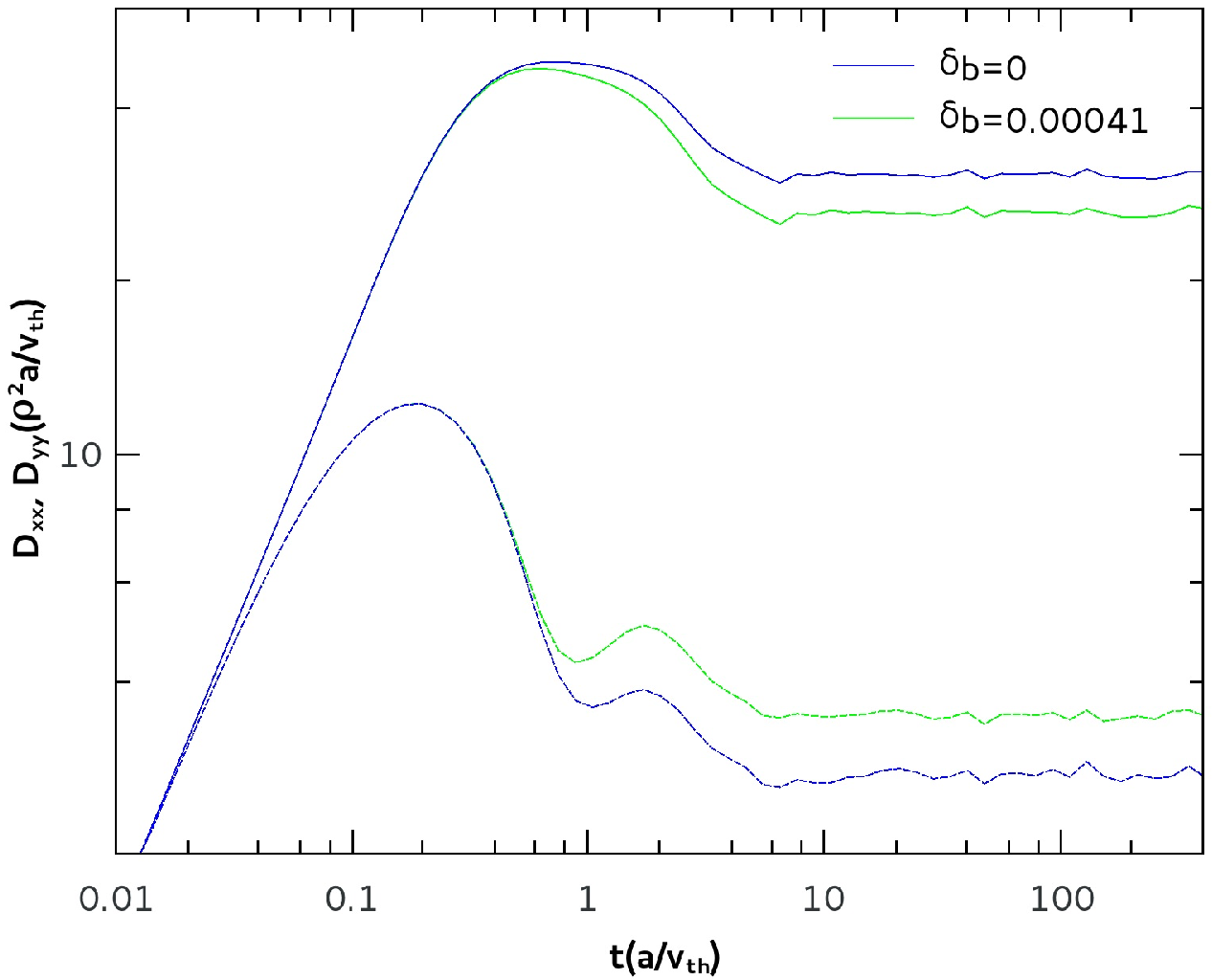}}
  \hspace{0.3cm}
  \subfloat[ $D_{xx}(t), D_{yy}(t)$ for different values of the parallel decorrelation length (blue lines $\lambda_z=0.1$, ciel $\lambda_z=0.5$, magenta $\lambda_z=1$, ) for a static potential $V_d=0$ and a magnetic field with $\delta b=0.0041$.]{\includegraphics[width=.48\linewidth, height=.38\linewidth]{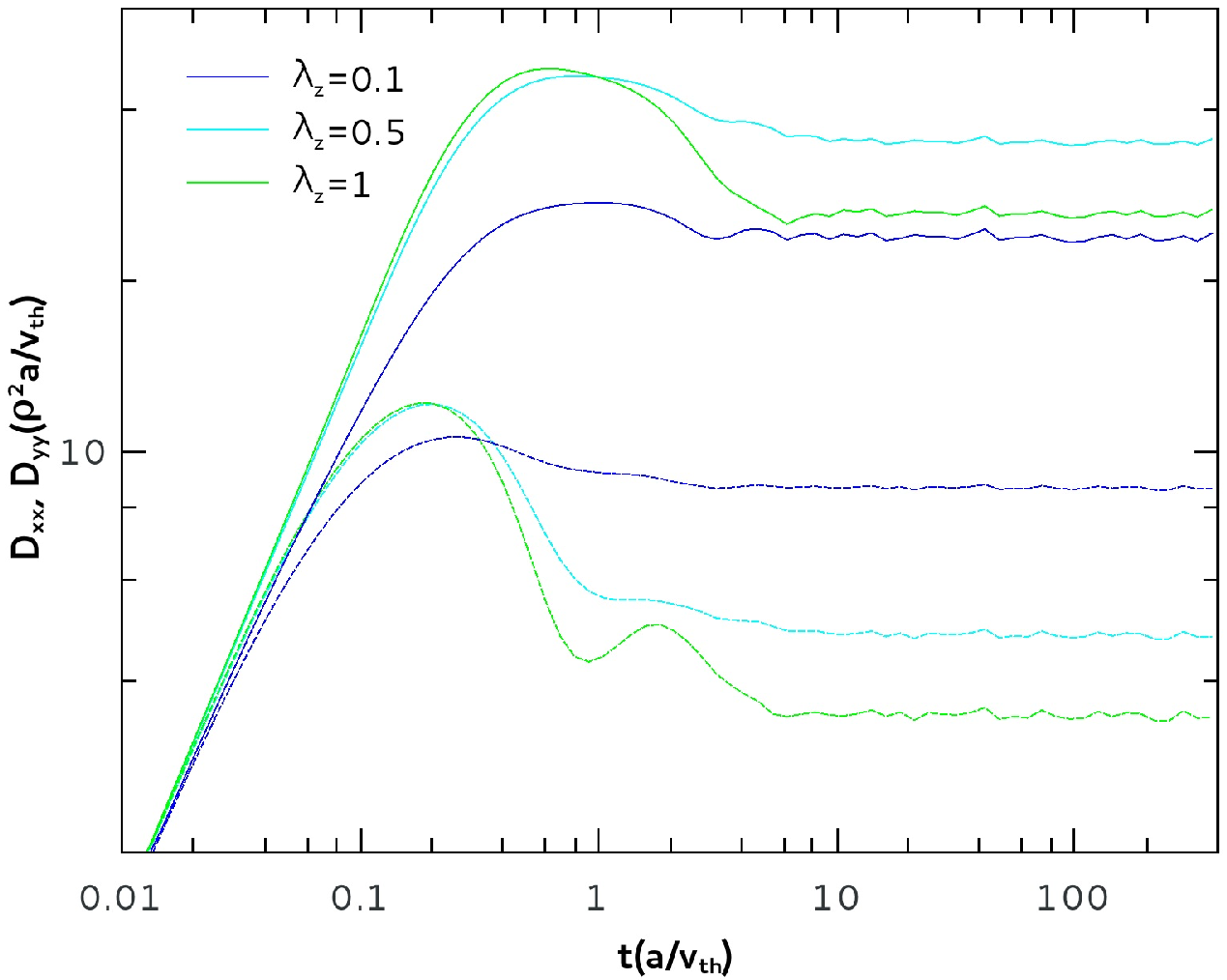}}   
\end{minipage}%
\caption{Running normalized diffusion coefficients for deuterium ions at the temperature $T=1keV$ in the nonlinear regime ($K_*$=10) in the limit of frozen turbulence ($\tau_c\to\infty$). Continuous lines correspond to $D_{xx}$ and dotted lines to $D_{yy}$. Other parameters are $\theta=0, \lambda_x=4,\lambda_y=2, k_2^0=1$.}
\label{taucinft}
\end{figure}

\begin{figure}
\begin{minipage}{.99\textwidth}
  \centering
  \subfloat[ $D_{xx}(t;\theta=0), D_{yy}(t;\theta=0)$ for $\tau_c\to\infty$  ]{\includegraphics[width=.48\linewidth, height=.38\linewidth]{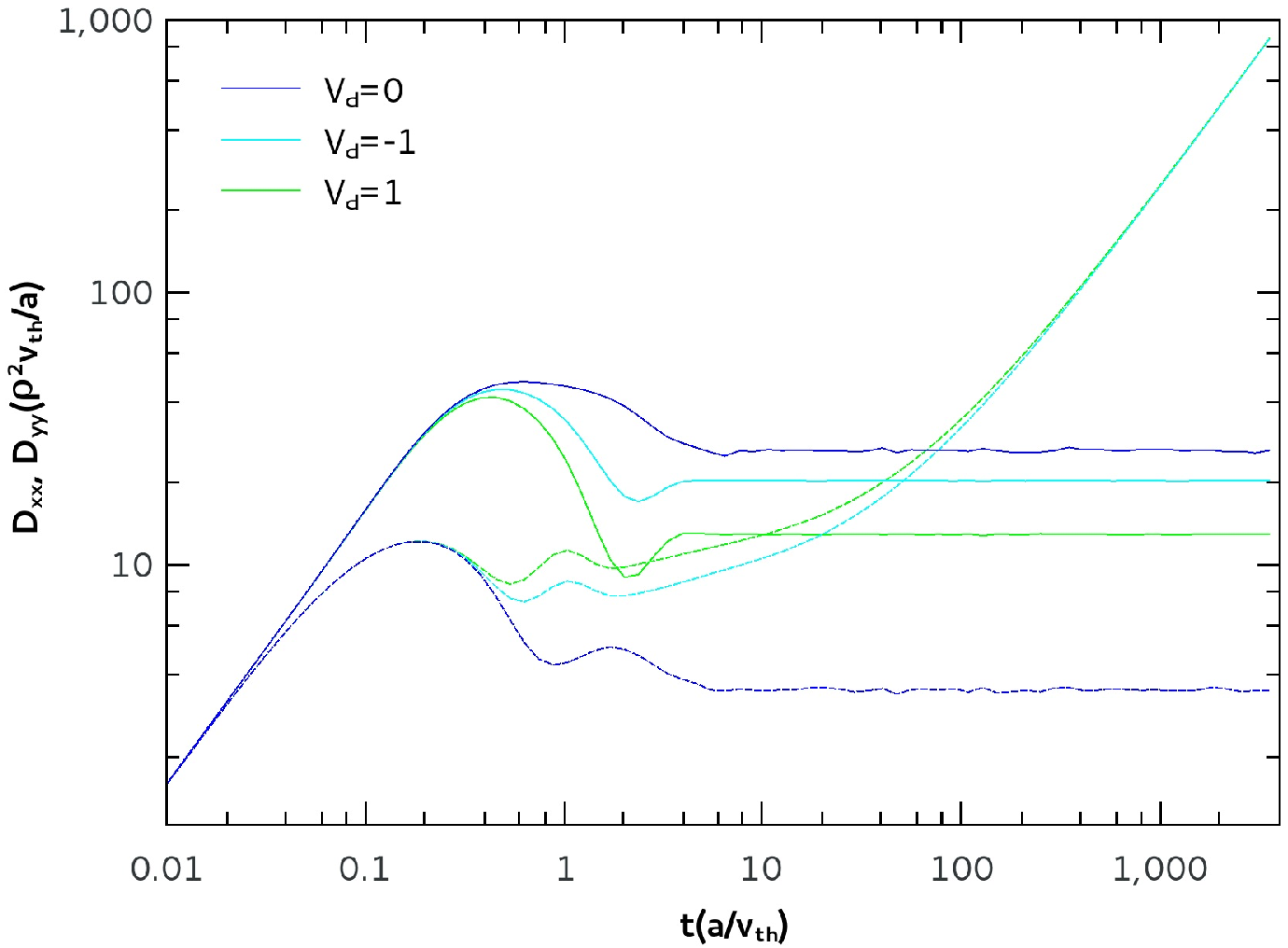}}
  \hspace{0.3cm}
  \subfloat[ $D_{xx}(t;\theta=0), D_{yy}(t;\theta=0)$ for $\tau_c=3$]{\includegraphics[width=.48\linewidth, height=.38\linewidth]{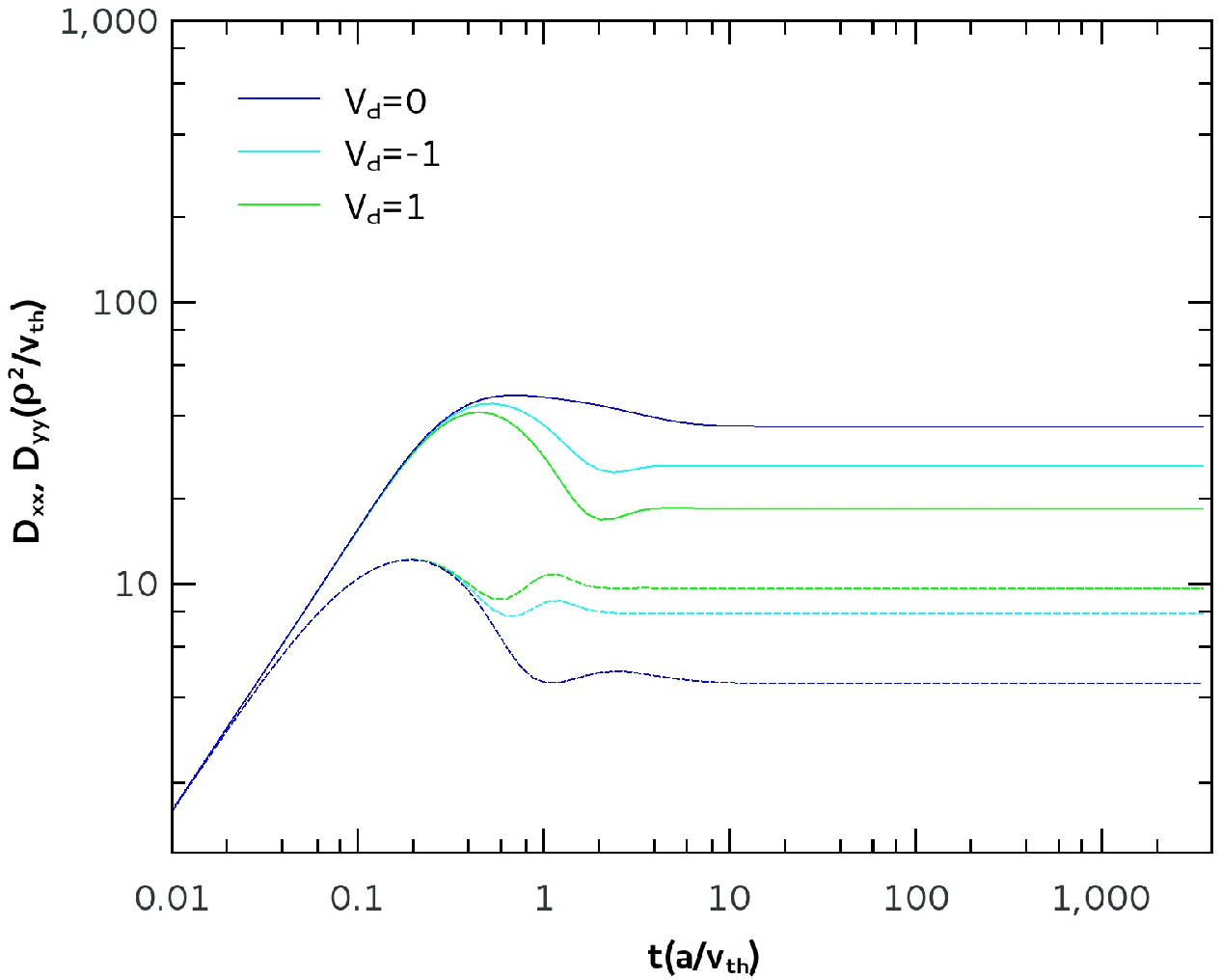}}   
\end{minipage}%
\caption{Running diffusion coefficients of deuterium ions at temperature $T=1keV$ in the nonlinear regime ($K_*$=10) in the limit of frozen turbulence ($\tau_c\to\infty$)(a) and for $\tau_c=3$(b) for different values of the modulus of the normalized diamagnetic velocity (blue $V_d=0$, $V_d=1$ oriented paralel (blue) and antiparallel (ciel) to the magnetic drift). Continuous lines correspond to $D_{xx}$ and dotted lines to $D_{yy}$. The other parameters are $\delta b=0.0041, \theta=0, \lambda_x=4,\lambda_y=2,\lambda_z=1, k_2^0=1$.}
\label{figlz1vdvar}
\end{figure}

\begin{figure}
\begin{minipage}{.99\textwidth}
  \centering
  \subfloat[ $D_{xx}(t;\theta), D_{yy}(t;\theta)$ for $\tau_c=3$  ]{\includegraphics[width=.48\linewidth, height=.38\linewidth]{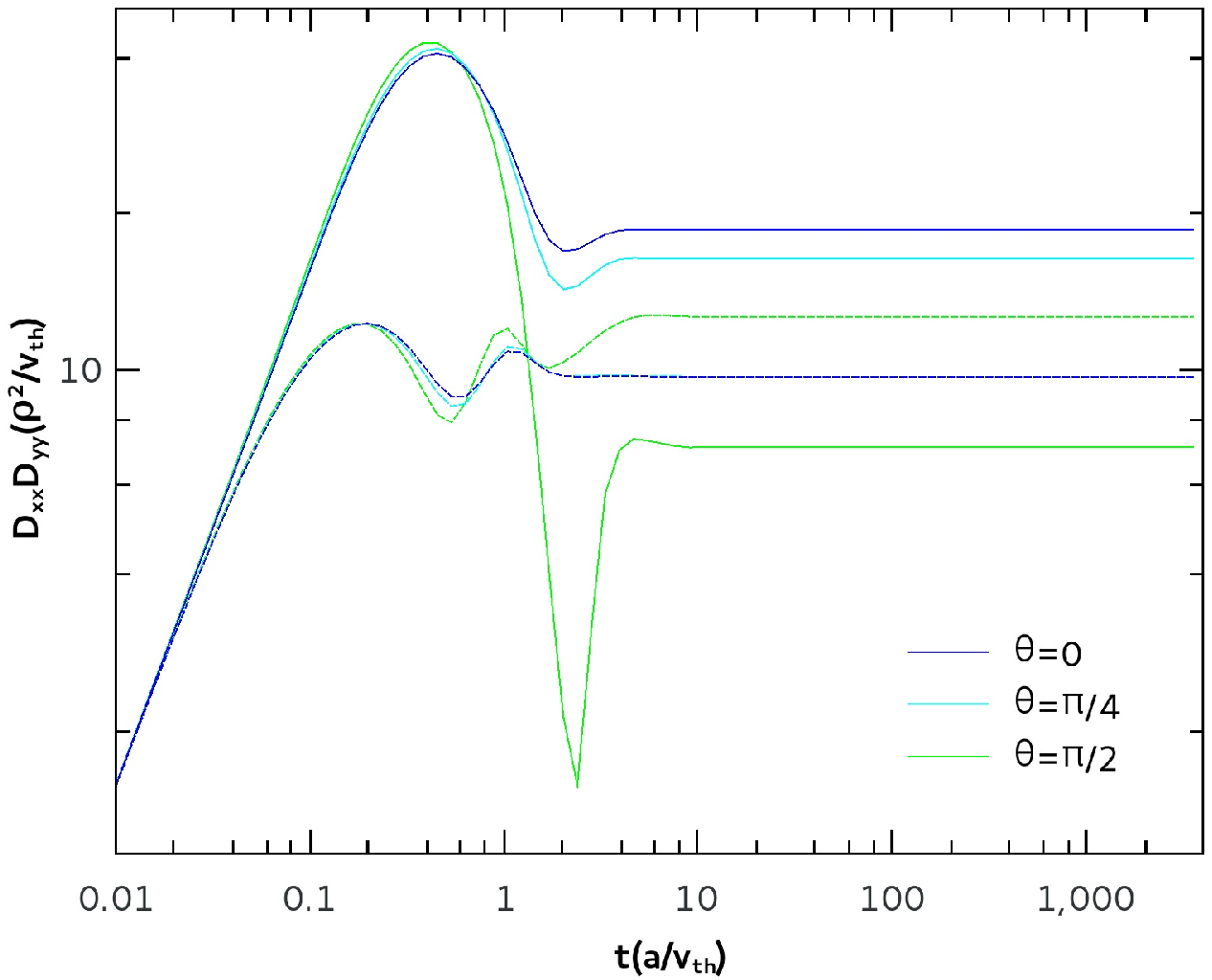}}
  \hspace{0.3cm}
  \subfloat[ $\lim_{t\to\infty}D_{xx}(t;\theta), D_{yy}(t;\theta)$ for $\tau_c=3$]{\includegraphics[width=.48\linewidth, height=.38\linewidth]{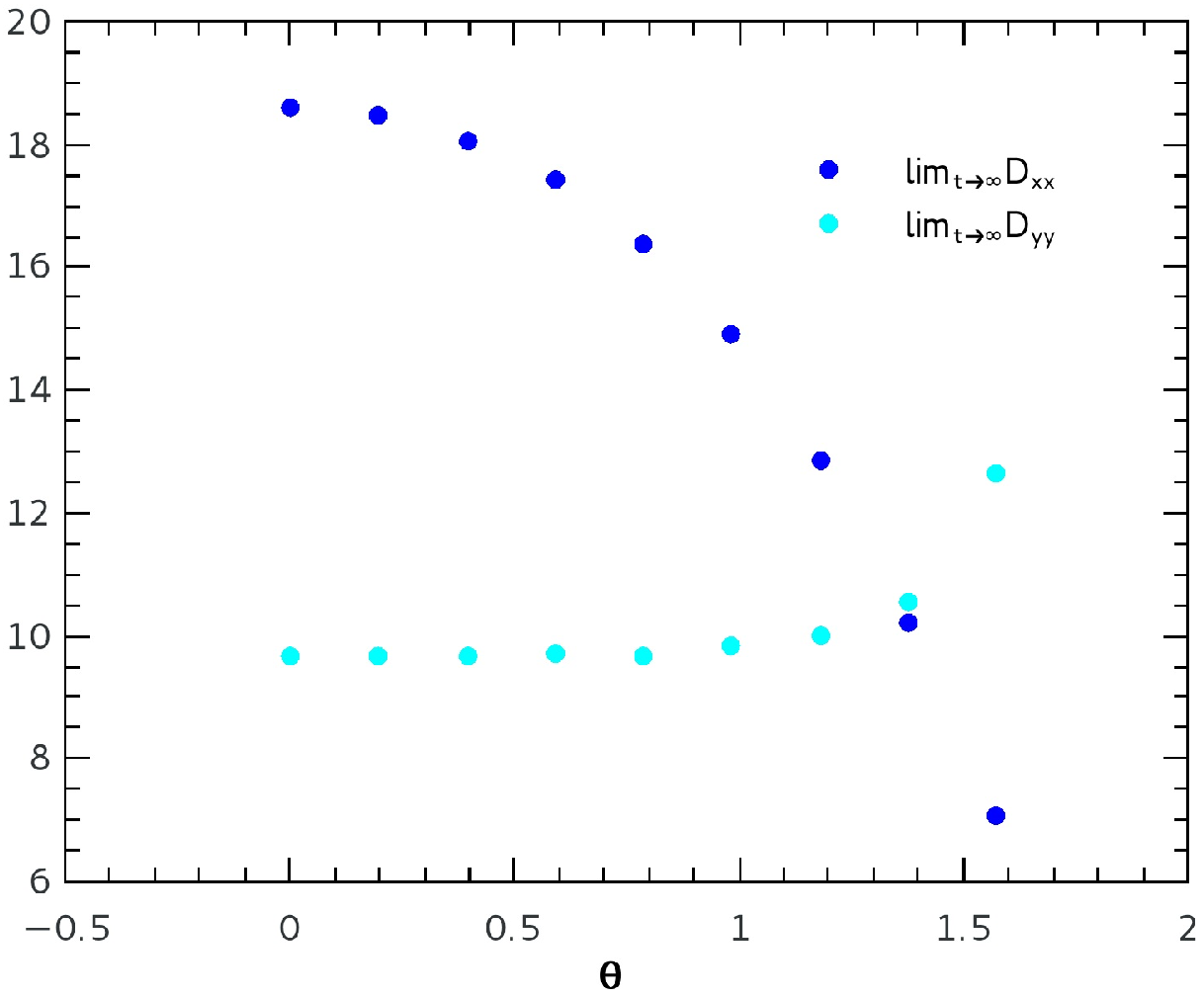}}   
\end{minipage}%
\caption{Running diffusion coefficients of deuterium ions at temperature $T=1keV$ in the nonlinear regime ($K_*$=10) at $\tau_c=3$ for different values of the pitch angle. Continuous lines correspond to $D_{xx}$ and dotted lines to $D_{yy}$. The other parameters are $\delta b=0.0041, \theta=0, \lambda_x=4,\lambda_y=2,\lambda_z=1, k_2^0=1, V_d=1$.}
\label{tetavar}
\end{figure}

The statistical model contains 9 physical parameters 
$K_*, \lambda_i, \tau_c, k_2^0, V_d, \delta_b$ and $\rho_*$ (or equivalent the temperature of the ions). These determine several characteristic times whose ordering determine the transport regimes. The decorrelation produced by the drift of the potential with the effective diamagnetic velocity induces a normalized diamagnetic time $\tau_d=\lambda_y/V_d$, the parallel motion of the particles induces a parallel decorrelation time  $\tau_\parallel=\lambda_z/v_\parallel\approx\lambda_z/v_{th}$, and the magnetic drift induces $\tau_b=\lambda_y\rho_*/\delta_b/(v_\parallel^2+v_\perp^2/2)$. The time of flight is $\tau_{fl}=\lambda_x/V_x+\lambda_y/V_y$ where $V_x$ and $V_y$ are the amplitudes of the $x$ and $y$ velocities respectively.

In the case of a static potential with no diamagnetic drift $V_d=0$ and in the limit of frozen turbulence $\tau_c\to\infty$ the decorrelation is provided by the magnetic drift and by the parallel motion of the particles. The characteristic time for the parallel motion is equal to the ratio between the parallel decorrelation length and the parallel velocity which is of the order of the thermal velocity. The normalized parallel decorrelation time will thus be equal to the normalized parallel decorrelation length $\lambda_z$, in units of the small radius of the tokamak, $\tau_\parallel/\tau_0=\lambda_z$ . In figure Fig. \ref{taucinft}(b) we plotted the running normalized diffusion coefficients for different values of the parallel decorrelation length. Trapping effects become visibile for $\lambda_z>1$. This offers a trapping condition $\tau_z>\tau_0$, since for smaller decorrelation times, particles moving radially can escape to the tokamak walls, without exploring the correlated region. As shown in \cite{vlad2006impurity, vlad2015nonlinear} the inhomogeneity of the magnetic field does not modify the shape of the running diffusion coefficient (see Fig. \ref{taucinft}(a)), its only effect being the cause of a direct transport. We thus choose a value of the normalized inverse gradient scale length of the magnetic field such that the ratio $\delta b/\rho_*=a/L_b\approx a/R\approx 0.3$ is relevant major to tokamak devices \cite{rebut1985joint}. This requires that deuterium ions at temperature $T=1keV$ will have $\delta b=0.00041$.

In figure Fig. \ref{figlz1vdvar}, we plotted the running diffusion coefficients for different values of the diamagnetic velocity. Since, as seen also in Fig. \ref{DTs}, at large times the particles move in the direction of the combined magnetic and diamagnetic drift, the transport in the poloidal direction increases dramatically. The antiparallel orientation of the diamagnetic velocity relative to the magnetic drift just decreases the total combined drift. In Fig. \ref{figlz1vdvar}(b) the finite $\tau_c$ decorrelates the particles from the contour lines of the potential, the transport becoming diffusive.

The contribution of the magnetic drift depends on the values of the pitch angle, see Fig. \ref{tetavar}. In the radial direction the diffusion coefficient decreases with the value of the pitch angle as expected, since for $\theta>0$ more particles are trapped in the magnetic field.

\section{Conclusions}\label{conclusions}

We studied the transport of low energy deuterium ions in a realistic model of tokamak microturbulence in the framework of the Decorelation Trajectory Method. The main decorrelation mechanisms are provided by the parallel motion of the ions, the diamagnetic velocity specific to the drift type turbulence and the neoclassical magnetic curvature and gradient drift due to the inhomogeneity of the magnetic field. These mechanisms induce several characteristic times whose ordering determine the transport regimes. In the limit of frozen turbulence we obtained that a neccessary condition for trapping is for the parallel decorrelation time to be greater or equal to the time needed for an ion moving radially to escape to the tokamak walls $\tau_z>\tau_0$. The magnetic drift and the drift of the potential with the diamagnetic velocity have similar effects on the trapping of the particles since they are equivalent with the existence of an additional potential $x*V_d+x*\frac{\delta b}{\rho_*}\left(v_\parallel^2+v_\perp^2/2\right)$ which modifies the contour lines of the total potential creating islands of closed lines that decrease in size as the drifts increases. The contribution of the magnetic drift depends on the value of the pitch angle and has a maximum in the radial direction at $\theta=0$.

\begin{acknowledgements}
This work was supported by the Romanian Ministry of
National Education under the contracts $1EU-4 WPJET1-RO_C$ and $PN 16 47 01 04$. The contract $1EU-4 WPJET1-RO_C$ is included in the Programme of Complementary Research in Fusion. The views presented here do not necessarily represent those of the European Commission. The author would also like to thank also M. Vlad and F. Spineanu for support and assistance in the development of this research project.
\end{acknowledgements}
\bibliographystyle{plain}
\bibliography{bibliography}

\end{document}